\documentclass[preprint, superscriptaddress, showpacs,preprintnumbers,amsmath,amssymb]{revtex4}
\usepackage{graphicx}

\begin{document}

\thispagestyle{empty}

\title{Constraints on corrections to Newtonian gravity
from two recent measurements of the Casimir interaction
between metallic surfaces}

\author{G.~L.~Klimchitskaya}
\affiliation{Central Astronomical Observatory
at Pulkovo of the Russian Academy of Sciences,
St.Petersburg, 196140, Russia
}
\author{U.~Mohideen}
\affiliation{Department of Physics and
Astronomy, University of California, Riverside, California 92521,
USA
}
\author{V.~M.~Mostepanenko}
\affiliation{Central Astronomical Observatory
at Pulkovo of the Russian Academy of Sciences,
St.Petersburg, 196140, Russia
}

\begin{abstract}
We obtain constraints on parameters
of the Yukawa-type corrections to Newton's gravitational law
from measurements of the gradient of the
Casimir force between surfaces coated with ferromagnetic metal Ni
and from measurements of the
Casimir force between Au-coated sinusoidally corrugated surfaces
at various angles between corrugations.
It is shown that constraints following from the experiment with
magnetic surfaces are slightly weaker than currently available
strongest constraints, but benefit from increased reliability and
independence of systematic effects.
The constraints derived from the experiment with corrugated
surfaces within the interaction region from 11.6 to 29.2\,nm
are stronger up to a factor of 4 than the strongest constraints
derived from other experiments.
The possibility of further strengthening of constraints on
non-Newtonian gravity by using the configurations with
corrugated boundaries is proposed.
\end{abstract}

\maketitle
\section{Introduction}

In the last few decades corrections to Newton's law of gravitation
and constraints on them have become the subject of considerable
study (see the monograph \cite{1} and reviews \cite{2,3,4,5}).
{}From the experimental standpoint, it is of most importance
that at separations between the test bodies below 0.1\,mm
Newton's law is not confirmed by measurements with
sufficient precision. Theoretically, many extensions of the
Standard Model of elementary particles and their interactions
predict corrections to the Newton law of power- and
Yukawa-type due to exchange of light and massless elementary
particles \cite{6,7,8,9,10}. On the other hand, similar
corrections are predicted by the extra-dimensional physics with
a low-energy compactification scale \cite{11,12,13,14,15}.
This makes the search for such corrections, or at least
constraining their parameters, interesting for the problems of
dark matter and unification of gravitation with other
fundamental interactions.

A lot of successful work has been done on constraining the
power- and Yukawa-type corrections to Newton's law of
gravitation from gravitational experiments of E\"{o}tvos-
and Cavendish-type \cite{1,2,3,4,16,17}. Although the most
strong constraints on the power-type corrections were obtained
in this way, it was found that the resulting constraints on
the Yukawa-type corrections become much weaker in the
interaction range below a few micrometers.
This is explained by the fact that at sufficiently small
separations between the test bodies the van der Waals \cite{18}
and Casimir force \cite{19} becomes the dominant background
force in place of gravitation. The two names belong to
a single force of quantum origin caused by the zero-point
and thermal fluctuations of the electromagnetic field.
They are
usually used at short and relatively large separations,
respectively, where the effects of relativistic retardation
are immaterial or, on the contrary, are influential and
should be taken into account.

The possibility to constrain corrections to Newton's law
from the van der Waals and Casimir force was proposed
long ago \cite{20,21} for the cases of Yukawa-type and
power-type corrections, respectively. At that time,
however, reasonably precise measurements
of the van der Waals and Casimir force were not available.
Things have changed during the last 15 years when
a lot of more precise experiments on measuring the Casimir
force between metallic, dielectric and semiconductor bodies
have been performed (see reviews \cite{22,23,24,25}).
The measure of agreement between the measurement data of
these experiments and theoretical description of the Casimir
force in the framework of the Lifshitz theory resulted in
the strengthening of previously known constraints on the
parameters of Yukawa-type corrections up to a factor of
$2.4\times 10^7$ \cite{5,19,26}. Using different experiments
on measurement of the Casimir force, the strongest constraints
on the corrections to Newton's law were obtained over a wide
interaction region from about 1\,nm to a few micrometers.
Note that for shorter interaction regions the strongest
constraints on the Yukawa-type corrections follow from precise
atomic physics \cite{27}, whereas starting from a few micrometers
the gravitational experiments \cite{1,2,3,4,16,17} remain the
most reliable source of such constraints.

In this paper we obtain constraints on the Yukawa-type
corrections to Newton's gravitational law from two recently
performed experiments on measuring the Casimir interaction by
means of an atomic force microscope (AFM).
Each of these experiments is highly significant, as compared with all
earlier measurements of the Casimir interaction.
In the first \cite{28} the dynamic AFM was used to
measure the gradient of the Casimir force between a plate and
a sphere both coated with the ferromagnetic metal Ni with no
spontaneous magnetization. As a result, the predictions of
the Lifshitz theory generalized for the case of magnetic
materials more than 40 years ago \cite{29} were experimentally
confirmed. The distinguished feature of the experiment with
two magnetic surfaces is also that it allows to shed light
on the role of some important systematic effects (see
Sec.~II for details) and, thus, remove any doubt in the
reliability of constraints obtained.

In the second experiment of our interest here \cite{30}
the static AFM was used to measure the Casimir force
between a plate and a sphere both with sinusoidally
corrugated surfaces coated with nonmagnetic metal Au.
The unusual feature of this experiment, as compared with
earlier performed experiments with corrugated surfaces,
is that the Casimir force was measured at various angles
between the longitudinal corrugations on both bodies.
This introduced into the problem an additional parameter
(the angle between corrugations) that can be chosen to
obtain the most strong constraints from the measure
of agreement between the experimental data and theory of
the Casimir force for corrugated surfaces based on the
derivative expansion \cite{31,32,33}.

The constraints on corrections to Newton's law obtained
by us from the experiment with magnetic surfaces are in
agreement with those obtained earlier \cite{34} for smooth
 Au surfaces by means of dynamic AFM \cite{35}, but
slightly weaker due to different densities of Au and Ni.
The advantage of constraints following from the experiment
with Ni surfaces is that they are not only fully justified
on their own, but add substantiation to the constraints
obtained from the experiments with nonmagnetic metal
surfaces. As to constraints obtained from the experiment
with corrugated surfaces, they are stronger up to a factor
of 4 than the most strong constraints reported so far
\cite{26,36,37} in the interaction region from 11.6 to
29.2\,nm.

The paper is organized as follows. In Sec.~II we present
the exact expression for the Yukawa-type interaction in
the experimental configuration of Ref.~\cite{28} and
derive the respective constraints on corrections to
Newton's gravitational law. The advantages of using
magnetic materials are elucidated.
Section~III is devoted to the experiment with corrugated
surfaces \cite{30}. Here, we derive an expression for the
Yukawa-type force in configurations with different angles
between corrugations. The obtained expression is used to
find the stronger constraints on corrections to
Newton's law. Some modifications in the setup are proposed
allowing further strengthening of the constraints in
configurations with corrugated surfaces. In Sec.~IV
the reader will find our conclusions and discussion.

\section{Constraints from the gradient of the Casimir
force {\protect \\} between two magnetic surfaces}

We begin with the standard parametrization of the
spin-independent Yukawa-type correction to Newton's
gravitational law \cite{1,2,3,4,5} (for spin-dependent
corrections see Refs.~\cite{38,39}).
The total gravitational potential
 between the two point-like masses
$m_1$ and $m_2$ spaced at a separation $r$ takes
the form
\begin{equation}
V(r)=V_N(r)+V_{\rm Yu}(r)=-\frac{Gm_1m_2}{r}\left(
1+\alpha e^{-r/\lambda}\right),
\label{eq1}
\end{equation}
\noindent
where $V_N(r)$ and $V_{\rm Yu}(r)$ are the Newtonian part
and the Yukawa-type correction, respectively.
Here, $G$ is the Newtonian
gravitational constant, and $\alpha$ and $\lambda$ are
the strength and interaction range of the Yukawa-type
correction. Similar to Ref.~\cite{40} it can be shown that
at separations below a few micrometers the Newtonian
gravitational force between the test bodies $V_1$ and $V_2$
in experiments under consideration is much smaller than the error
in measurements of the Casimir force. Because of this, in all
subsequent calculations the Newtonian potential can be
neglected, and the Yukawa-type addition to it is considered on
the background of the measured Casimir force. Then the
gravitational force acting between the test bodies at short
separations can be obtained by the integration of the
Yukawa-type correction $V_{\rm Yu}(r)$ defined in Eq.~(\ref{eq1})
over the volumes of both bodies
\begin{equation}
V_{\rm Yu}(a)=-G\alpha\int_{V_1}d^3r_1
\rho_1(\mbox{\boldmath$r$}_1)\int_{V_2}d^3r_2
\rho_2(\mbox{\boldmath$r$}_2)
\frac{e^{-|{\scriptsize{\mbox{\boldmath$r$}_1-
\mbox{\boldmath$r$}_2}}|/\lambda}}{|\mbox{\boldmath$r$}_1-
\mbox{\boldmath$r$}_2|}.
\label{eq2}
\end{equation}
\noindent
Here, $a$ is the closest separation between the test bodies
and $\rho_1(\mbox{\boldmath$r$}_1)$ and
$\rho_2(\mbox{\boldmath$r$}_2)$ are the respective mass densities
(note that $\rho_1$ and $\rho_2$ are not constant because in the
experiments used below each test body consists of several
homogeneous layers of different densities). The gravitational
force due to the Yukawa-type correction and its gradient
are given by
\begin{equation}
F_{\rm Yu}(a)=-\frac{\partial V_{\rm Yu}(a)}{\partial a},
\qquad
\frac{\partial F_{\rm Yu}(a)}{\partial a}
=-\frac{\partial^2 V_{\rm Yu}(a)}{\partial a^2}.
\label{eq3}
\end{equation}

In the experiment \cite{28} the gradient of the Casimir force was
measured between a Ni-coated plate and a Ni-coated hollow
microsphere
attached to the tip of an AFM cantilever operated in the dynamic
regime \cite{19,22}.
The silicon plate ($V_1$)
of a few millimeter diameter and thickness
can be considered as having an infinitely large area and an
infinitely large thickness when we have to deal with the
submicrometer region of $\lambda$.
The density of Si is
$\rho_{\,\rm Si}=2.33\times 10^3\,\mbox{kg/m}^{3}$.
For technological purposes the Si plate was coated first with a
layer of Cr having a thickness
$\Delta_{\rm Cr}^{\!(1)}=10\,$nm
and density
$\rho_{\,\rm Cr}=7.15\times 10^3\,\mbox{kg/m}^{3}$
and then with a
layer of Al having a thickness
$\Delta_{\rm Al}^{\!(1)}=40\,$nm
and density
$\rho_{\,\rm Al}=2.7\times 10^3\,\mbox{kg/m}^{3}$.
Finally the plate was coated with an outer layer of
magnetic metal Ni with a thickness
$\Delta_{\rm Ni}^{\!(1)}=250\,$nm
and density
$\rho_{\,\rm Ni}=8.9\times 10^3\,\mbox{kg/m}^{3}$.
The hollow microsphere ($V_2$) was made of glass
with density
$\rho_{g}=2.5\times 10^3\,\mbox{kg/m}^{3}$.
The thickness of the spherical envelope was
$\Delta_{g}^{\!(2)}=5\,\mu$m.
The sphere was also coated with successive layers of Cr, Al and
Ni having the thicknesses
$\Delta_{\rm Cr}^{\!(2)}=\Delta_{\rm Cr}^{\!(1)}$,
$\Delta_{\rm Al}^{\!(2)}=\Delta_{\rm Al}^{\!(1)}$,
and
$\Delta_{\rm Ni}^{\!(2)}=210\,$nm.
The outer radius of the sphere with all the coatings included is
$R=61.7\,\mu$m.

The exact integration over the volumes of a plate and a sphere in
Eq.(\ref{eq2}) with account of their layer structure can be
performed like in Ref.~\cite{41}. Then, substituting the obtained
result in Eq.~(\ref{eq3}), we arrive at
\begin{equation}
\frac{\partial F_{\rm Yu}(a)}{\partial a}=
4\pi^2G\alpha\lambda^2e^{-a/\lambda}X^{(1)}(\lambda)X^{(2)}(\lambda),
\label{eq4}
\end{equation}
\noindent
where the following notations are introduced:
\begin{eqnarray}
&&
X^{(1)}(\lambda)=\rho_{\,\rm Ni}-(\rho_{\,\rm Ni}-\rho_{\,\rm Al})
e^{-\Delta_{\rm Ni}^{\!(1)}/\lambda}
\nonumber \\
&&~~~~
-(\rho_{\,\rm Al}-\rho_{\,\rm Cr})
e^{-(\Delta_{\rm Ni}^{\!(1)}+\Delta_{\rm Al}^{\!(1)})/\lambda}
\nonumber \\
&&~~~~
-(\rho_{\,\rm Al}-\rho_{\,\rm Si})
e^{-(\Delta_{\rm Ni}^{\!(1)}+\Delta_{\rm Al}^{\!(1)}
+\Delta_{\rm Cr}^{\!(1)})/\lambda},
\label{eq5}\\
&&
X^{(2)}(\lambda)=\rho_{\,\rm Ni}\Phi(R,\lambda)-
(\rho_{\,\rm Ni}-\rho_{\,\rm Al})\Phi(R-\Delta_{\rm Ni}^{\!(2)},\lambda)
e^{-\Delta_{\rm Ni}^{\!(2)}/\lambda}
\nonumber \\
&&~~
-(\rho_{\,\rm Al}-\rho_{\,\rm Cr})\Phi(R-\Delta_{\rm Ni}^{\!(2)}-\Delta_{\rm Al}^{\!(2)},\lambda)
e^{-(\Delta_{\rm Ni}^{\!(2)}+\Delta_{\rm Al}^{\!(2)})/\lambda}
\nonumber \\
&&~~
-(\rho_{\,\rm Cr}-\rho_{g})
\Phi(R-\Delta_{\rm Ni}^{\!(2)}-\Delta_{\rm Al}^{\!(2)}-\Delta_{\rm Cr}^{\!(2)},\lambda)
e^{-(\Delta_{\rm Ni}^{\!(2)}+\Delta_{\rm Al}^{\!(2)}+\Delta_{\rm Cr}^{\!(2)})/\lambda}
\nonumber \\
&&~~
-\rho_{g}\Phi(R-\Delta_{\rm Ni}^{\!(2)}-\Delta_{\rm Al}^{\!(2)}
-\Delta_{\rm Cr}^{\!(2)}-\Delta_{g}^{\!(2)},\lambda)
e^{-(\Delta_{\rm Ni}^{\!(2)}+
\Delta_{\rm Al}^{\!(2)}+\Delta_{\rm Cr}^{\!(2)}+\Delta_g^{\!(2)})/\lambda},
\nonumber
\end{eqnarray}
\noindent
and the following notation is introduced
\begin{equation}
\Phi(r,\lambda)=r-\lambda+(r+\lambda)e^{-2r/\lambda}.
\label{eq6}
\end{equation}

The constraints on the parameters ($\lambda,\alpha$), which are
often referred to as the parameters of
{\it non-Newtonian gravity},
can be obtained from the comparison between  the measurement data
for the gradient of the Casimir force $F_{C}^{\prime}(a)$ and
respective theory. In Ref.~\cite{28} it was found that within the
entire separation region from 223 to 550\,nm there is an
excellent agreement between the data and theoretical predictions
of the Lifshitz theory of the van der Waals and Casimir force
\cite{18,19} with omitted relaxation properties of conduction
electrons (the so-called {\it plasma model approach}).
The  predictions of the Lifshitz theory
 with included relaxation properties of free charge carriers
(the so-called {\it Drude model approach}) were excluded by the
measurement data at a 95\% confidence level within the separation
region from 223 to 350\,nm (in the end of this section we provide
a brief discussion of different approaches to the Lifshitz theory
which is essential for obtaining constraints on non-Newtonian
gravity). The measure of agreement with the adequate theory is
characterized by the total experimental error
$\Delta_{F_C^{\prime}}(a)$ in the measured gradient of the Casimir
force determined at a 67\% confidence level \cite{28}.
Keeping in mind that within the limits of this error no
additional interaction of Yukawa-type was observed, the
constraints
on the parameters $\lambda$ and $\alpha$ can be obtained
from the inequality
\begin{equation}
\left|\frac{\partial F_{\rm Yu}(a)}{\partial a}\right|
\leq\Delta_{F_C^{\prime}}(a).
\label{eq7}
\end{equation}

\begin{figure}[b]
\vspace*{-12cm}
\centerline{\hspace*{3cm}
\includegraphics{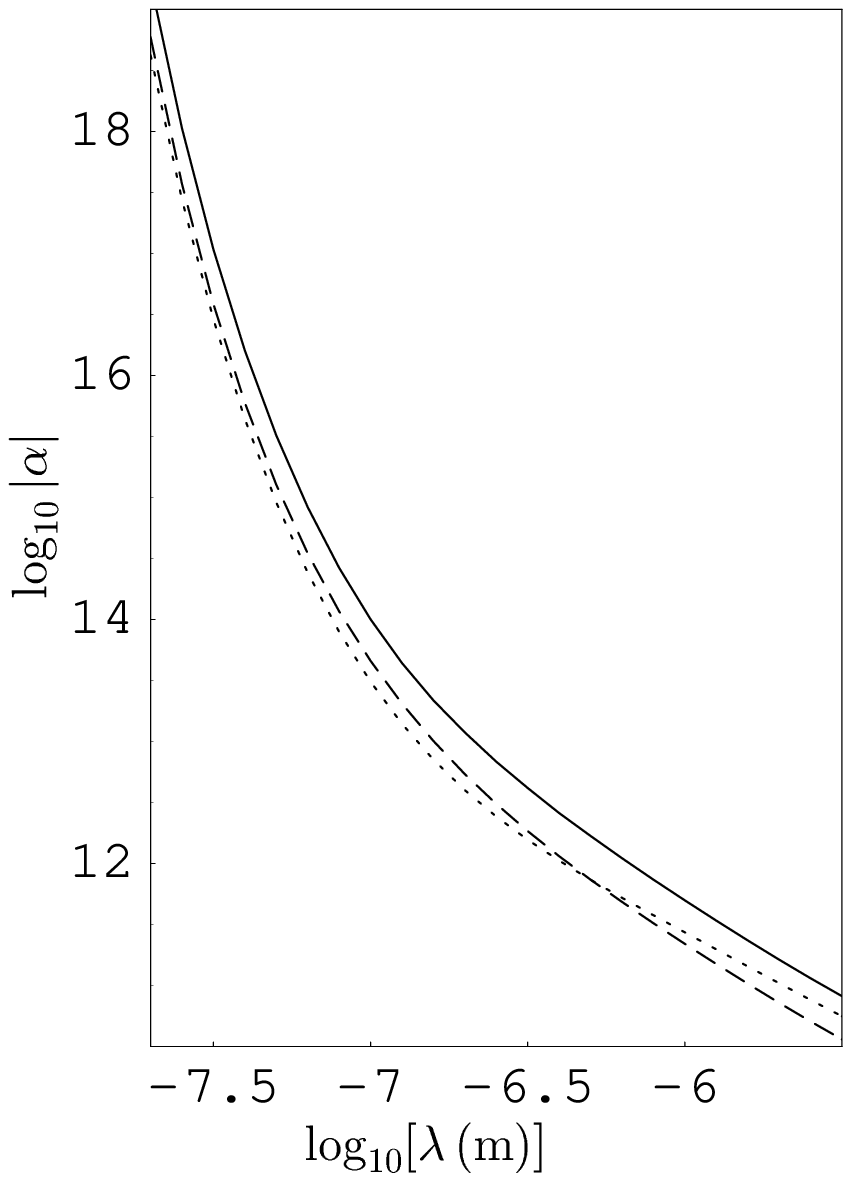}
}
\vspace*{-7cm}
\caption{\label{fg1}
Constraints on the parameters of Yukawa-type corrections to
Newton's gravitational law obtained in this work from
measurement of the gradient of the Casimir force between
two Ni surfaces (solid line), between two Au surfaces
(dashed line) and between an Au  and a Ni surfaces
(dotted line). Here and in Figs.~\ref{fg2},\,\ref{fg4}
the regions of $(\lambda,\alpha)$ plane below each line
are allowed and above each line are prohibited (see text
for further discussion).
}
\end{figure}

We have substituted Eqs.~(\ref{eq4})--(\ref{eq6}) in
Eq.~(\ref{eq7})
and analyzed the resulting inequality at different separations.
It was found that for
$\lambda\lesssim 200\,$nm
the strongest constraints are determined at the shortest
separation
$a=223\,$nm where $\Delta_{F_C^{\prime}}=1.2\,\mu$N/m \cite{28}.
For
$200\,\mbox{nm}\lesssim\lambda\lesssim 315\,$nm
and
$315\,\mbox{nm}\lesssim\lambda\lesssim 630\,$nm
the strongest constraints follow at $a=250\,$ and 300\,nm,
respectively (with respective $\Delta_{F_C^{\prime}}=1.05$
and $0.89\,\mu$N/m). Finally, at
$\lambda>630\,$nm
the strongest constraints are obtained at $a=350\,$nm
($\Delta_{F_C^{\prime}}=0.81\,\mu$N/m).
The resulting constraints are shown by the solid line in
Fig.~\ref{fg1}. Here and below the region of
($\lambda,\alpha$) plane above each line is prohibited and
below is allowed by the results of respective experiment.
In the same figure by the dashed line we show the constraints
obtained in Ref.~\cite{34} from measurements of the gradient
of the Casimir force between two Au-coated surfaces by means
of dynamic AFM \cite{35}. The dotted line shows the constraints
obtained \cite{34} from the experiment on measuring gradient
of the Casimir force between an Au-coated sphere and a Ni-coated
plate \cite{42} using the same setup.
As can be seen in Fig.~\ref{fg1}, the constraints indicated by
the solid line are slightly weaker than those shown by the
dashed and dotted lines. This is caused by the fact that
density of Ni is smaller than density of Au and by different
experimental errors. Note also that our constraints shown by the
solid line can be obtained in a simpler way by using the
proximity force approximation \cite{19,22}
\begin{equation}
F_{\rm Yu}(a)=2\pi R E_{\rm Yu}(a),
\label{eq7a}
\end{equation}
\noindent
to calculate the
gradient of the Yukawa-type force, where $E_{\rm Yu}(a)$ is the
energy per unit area of Yukawa-type interaction between two
plane-parallel plates having the same layer structure as our
test bodies.
According to the results of
Refs.~\cite{41,43}, this is possible
under the conditions
\begin{equation}
\frac{\lambda}{R}\ll 1, \qquad
\frac{\Delta_{\rm Au}^{\!(2)}+\Delta_{\rm Al}^{\!(2)}+
\Delta_{\rm Cr}^{\!(2)}+\Delta_g^{\!(2)}}{R}\ll 1,
\label{eq8}
\end{equation}
\noindent
which are satisfied in our experimental configuration
with a wide safety margin.
In this case the function $\Phi(r,\lambda)$
with any argument $r$ can be approximately
replaced with $R$.

\begin{figure}[t]
\vspace*{-16cm}
\centerline{\hspace*{3cm}
\includegraphics{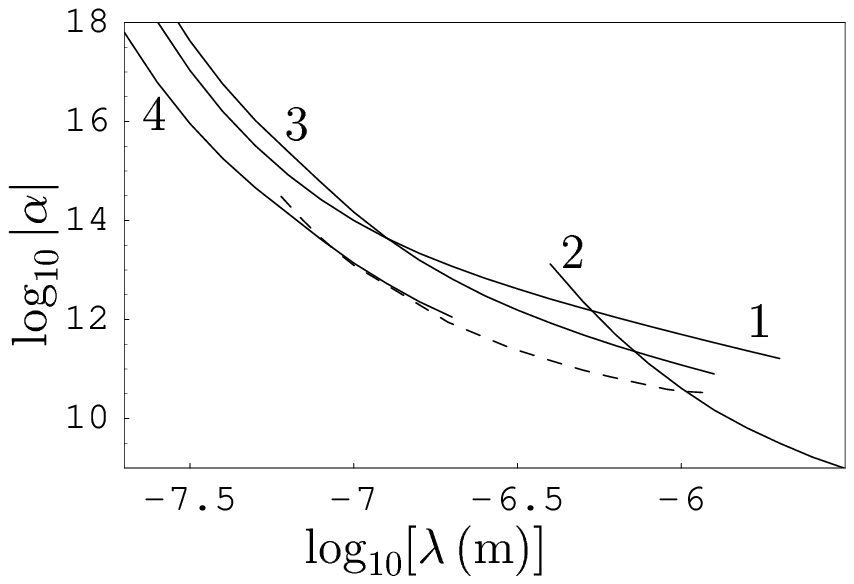}
}
\vspace*{-7cm}
\caption{\label{fg2}
Constraints on the parameters of Yukawa-type corrections to
Newton's gravitational law obtained in this work
(solid line 1), in Ref.~\cite{26} from measurements
of the thermal Casimir-Polder force \cite{44} (solid line 2),
in Ref.~\cite{45} from measurements of the gradient of the
Casimir force between metallic and corrugated semiconductor
surfaces \cite{46} (solid line 3), in Refs.~\cite{36,37}
from measurements of the gradient of the Casimir force
between two metallic surfaces (solid line 4), and in
Ref.~\cite{47} from the Casimir-less experiment
(dashed line).
}
\end{figure}

It would be interesting also to compare the constraints on
non-Newtonian gravity obtained here from the experiment with two
Ni surfaces (solid line in Fig.~\ref{fg1}) with the strongest
constraints obtained so far using the alternative setups.
For this purpose in Fig.~\ref{fg2} we reproduce the solid line
of Fig.~\ref{fg1} as the solid line 1. The solid line 2
in Fig.~\ref{fg2} was obtained \cite{26} from measurements of
the thermal Casimir-Polder force between ${}^{87}$Pb atoms
belonging to the Bose-Einstein condensate and a SiO${}_2$
plate \cite{44}, and the solid line 3 was obtained \cite{45}
from measurements of gradient of the Casimir force
between an Au sphere and a rectangular corrugated semiconductor
(Si) plate by means of a micromachined oscillator \cite{46}.
Next, the solid line 4 in Fig.~\ref{fg2} was found from an
effective measurement of the Casimir pressure between two
parallel Au plates by means of a micromachined oscillator
\cite{36,37}, and the dashed line was obtained from the
Casimir-less experiment \cite{47}. As can be seen
in Fig.~\ref{fg2}, various constraints obtained using
quite different setups are consistent with the constraints
of line 1 obtained from the most recent experiment with
two magnetic surfaces.

In the end of this section it is pertinent to note that the
experiment with two magnetic surfaces \cite{28} plays the key
role in the test of validity of the Lifshitz theory.
Keeping in mind that constraints on non-Newtonian gravity
are derived from the measure of agreement between the
measurement data and theory, this experiment is also
important to validate the reliability of constraints obtained.
As mentioned above, the Lifshitz theory is in agreement
with the plasma model approach to the Casimir force, which
disregards the relaxation properties of free charge carriers,
and excludes the Drude model approach taking these properties
into account (see the experiments of Refs.~\cite{35,36,37}
and earlier experiments reviewed in Refs.~\cite{19,22}).
This is against expectations of many and gave rise
to the search of some systematic effects which could reverse
the situation. After several unsuccessful attempts (see
Ref.~\cite{48} for a review) the influence of large surface
patches was selected as the most probable systematic effect
which could bring the data in agreement with the Drude model
approach \cite{49}. In two experiments on measuring the
Casimir force between Au surfaces \cite{50,51} hypothetical
large patches were described by models with free
fitting parameters and used in respective fitting
procedures. In these experiments, which are not independent
measurements of the Casimir force, an agreement of the data
with the Drude model approach has been claimed (see
Refs.~\cite{52,53,54,55,56} for a critical discussion).

The crucial point to underline here is that for nonmagnetic
metals the Drude model approach leads to smaller gradients of
the Casimir force than the plasma model approach
\cite{19,22,35,36,37}. Thus, the effect of large patch
potentials (which leads to an attraction similar to the
Casimir force) is added to the predictions of the Drude model
approach and might make the total theoretical force compatible
with the measurement data \cite{49}. By contrast, for two
magnetic metals the Drude model approach leads to larger
gradients of the Casimir force than the plasma model approach
\cite{28,57,58}. Thus, if the effect of patches were important in
this case, it would further increase the disagreement between the
predictions of the Drude model approach and the measurement data
observed in Ref.~\cite{28}. This confirms that surface patches
do not play an important role in precise experiments on
measuring the Casimir force in accordance with the model of
patches \cite{59} predicting a negligibly small effect
from patches \cite{19,22}. Recently the patches on Au
samples used in measurements of the Casimir force were
investigated by means of Kelvin probe microscopy \cite{60}.
The force originating from them was found to be too small
to affect the conclusions following from precise measurements
of the Casimir force.
It is the matter of fact that the experimental data of all
independent measurements of the Casimir interaction between
both nonmagnetic and magnetic metals are in excellent
agreement with the predictions of the Lifshitz theory
combined with the plasma model approach and exclude the Drude
model approach. Although the fundamental reasons behind this
fact have not yet been finally understood, the constraints
on non-Newtonian gravity obtained on this basis can be
already considered as reliable enough.

\section{Constraints from the Casimir force between two
corrugated surfaces}

In Sec.~II we have used the most recent measurement of the Casimir
interaction where the material dependence played a major role in
theory-experiment comparison.
Another recent experiment \cite{30} is of quite a different nature.
In Ref.~\cite{30} the normal Casimir force acting perpendicular
to the surface was measured between the sinusoidally corrugated
surfaces of a sphere and a plate. The corrugated boundary
surfaces have long been used in measurements of the Casimir force
(see Refs.~\cite{22,23} for a review). For example, the normal
Casimir force between a rectangular corrugated semiconductor (Si)
plate and a smooth Au sphere has been measured by means of a
micromachined oscillator and compared with theory based on the
exact scattering approach \cite{46}. The obtained constraints
on non-Newtonian gravity are discussed in Sec.~II (see solid
line 3 in Fig.~\ref{fg2}). A further example is the lateral
Casimir force between  a sinusoidally corrugated plate and
a sinusoidally corrugated sphere, both coated with Au, which was
measured and compared with exact theory in Refs.~\cite{61,62}.
This experiment resulted in the maximum strengthening of
constraints
on non-Newtonian gravity from the Casimir effect by a factor of
$2.4\times 10^{7}$ discussed in Sec.~I.
In experiments with corrugated surfaces the nontrivial geometry
plays a major role in the theory-experiment comparison  whereas
different approaches to the description of material properties
cannot be differentiated due to the lower experimental precision.

The specific feature of the experiment of Ref.~\cite{30} is that
the normal Casimir force between a sinusoidally corrugated
Au-coated plate and a sinusoidally corrugated Au-coated sphere
was measured at various angles between corrugations using an
AFM. The plate in this experiment is the diffraction grating
with uniaxial sinusoidal corrugations of period
$\Lambda=570.5\,$nm and amplitude $A_1=40.2\,$nm.
The grating was made of hard epoxy with density
$\rho_e=1.08\times 10^3\,\mbox{kg/m}^3$ and coated with an Au
layer of thickness $\Delta_{\rm Au}^{\!(1)}=300\,$nm.
The corrugated plate was used as a template for the pressure
imprinting of the corrugations on the bottom surface of a sphere.
The polystyrene sphere has a density
$\rho_p=1.06\times 10^3\,\mbox{kg/m}^3$.
It was coated with a layer of Cr of thickness
$\Delta_{\rm Cr}^{\!(2)}=10\,$nm, then with a layer of Al of thickness
$\Delta_{\rm Al}^{\!(2)}=20\,$nm and finally
with a layer of Au of thickness
$\Delta_{\rm Au}^{\!(2)}=110\,$nm.
The outer radius of the coated sphere is $R=99.6\,\mu$m.
The imprinted corrugations on the sphere have the same period as
on the plate and the amplitude $A_2=14.6\,$nm.
The size of an imprint area was measured to be
$L_x\approx L_y\approx 14\,\mu$m, i.e., it is much larger
than $\Lambda$. In Ref.~\cite{30} the Casimir force between the
sphere and the plate was measured at the following angles between
the axes of corrugations on both bodies:
$\theta=0^{\circ}$, $1.2^{\circ}$, $1.8^{\circ}$, and
$2.4^{\circ}$.

Now we calculate the Yukawa-type force in the experimental
configuration of Ref.~\cite{30}. For this purpose we first
consider
the Yukawa-type energy per unit area in the configuration
of two plane-parallel plates spaced at a separation $a$ having the
same layer structure as a plate and a sphere in the
experiment. The result is \cite{41}
\begin{equation}
E_{\rm Yu}(a)=
-2\pi G\alpha\lambda^3e^{-a/\lambda}X^{(1)}(\lambda)X^{(2)}(\lambda),
\label{eq9}
\end{equation}
\noindent
where now
\begin{eqnarray}
&&
X^{(1)}(\lambda)=\rho_{\,\rm Au}-
(\rho_{\,\rm Au}-\rho_{e})
e^{-\Delta_{\rm Au}^{\!(1)}/\lambda},
\label{eq10} \\
&&
X^{(2)}(\lambda)=\rho_{\,\rm Au}-
(\rho_{\,\rm Au}-\rho_{\,\rm Al})
e^{-\Delta_{\rm Au}^{\!(2)}/\lambda}
\nonumber \\
&&~~
-(\rho_{\,\rm Al}-\rho_{\,\rm Cr})
e^{-(\Delta_{\rm Au}^{\!(2)}+\Delta_{\rm Al}^{\!(2)})/\lambda}
\nonumber \\
&&~~
-(\rho_{\,\rm Cr}-\rho_{p})
e^{-(\Delta_{\rm Au}^{\!(2)}+\Delta_{\rm Al}^{\!(2)}+
\Delta_{\rm Cr}^{\!(2)})/\lambda}.
\nonumber
\end{eqnarray}
\noindent
Next, we introduce corrugations at an angle $\theta$ on the
parallel plates and find their effect by means of the geometrical
averaging \cite{19,22}
\begin{equation}
E_{\rm Yu}^{\rm corr}(a)=\frac{1}{L_xL_y}
\int_{-L_x/2}^{L_x/2}\!\!dx
\int_{-L_y/2}^{L_y/2}\!\!dy\,E_{\rm Yu}\Big(z(a,x,y)\Big).
\label{eq11}
\end{equation}
\noindent
Here, $E_{\rm Yu}$ is the energy per unit area defined in
Eq.~(\ref{eq9})
calculated at different separations $z$ between the corrugated
plates which are assumed parallel to the ($x,y$) plane
\begin{equation}
z(a,x,y)=a+A_1\cos\frac{2\pi x}{\Lambda}-
A_2\cos\frac{2\pi x^{\prime}}{\Lambda}.
\label{eq12}
\end{equation}
\noindent
Note that there is no phase shift between the corrugations on
both plates, so that
$x^{\prime}=x\cos\theta-y\sin\theta$.

Finally, to obtain the Yukawa-type force between a corrugated
plate and a corrugated sphere, we apply the proximity force
approximation (\ref{eq7a}) taking into account different radii
of separate spherical layers. After an easy calculation
using Eqs.~(\ref{eq9})--(\ref{eq12}),
the  Yukawa-type force between
a corrugated plate and a corrugated sphere takes the form
\begin{equation}
 F_{\rm Yu}^{\rm corr}(a)=
-4\pi^2G\alpha\lambda^3e^{-a/\lambda}X^{(1)}(\lambda)
\tilde{X}^{(2)}(\lambda)X(\lambda,\theta),
\label{eq13}
\end{equation}
\noindent
where
\begin{eqnarray}
&&
\tilde{X}^{(2)}(\lambda)=R\rho_{\,\rm Au}-
(\rho_{\,\rm Au}-\rho_{\,\rm Al})(R-\Delta_{\rm Au}^{\!(2)})
e^{-\Delta_{\rm Au}^{\!(2)}/\lambda}
\nonumber \\
&&~~
-(\rho_{\,\rm Al}-\rho_{\,\rm Cr})
(R-\Delta_{\rm Au}^{\!(2)}-\Delta_{\rm Al}^{\!(2)})
e^{-(\Delta_{\rm Au}^{\!(2)}+\Delta_{\rm Al}^{\!(2)})/\lambda}
\label{eq14} \\
&&~~
-(\rho_{\,\rm Cr}-\rho_{p})
(R-\Delta_{\rm Au}^{\!(2)}-\Delta_{\rm Al}^{\!(2)}-\Delta_{\rm Cr}^{\!(2)})
e^{-(\Delta_{\rm Au}^{\!(2)}+\Delta_{\rm Al}^{\!(2)}+\Delta_{\rm Cr}^{\!(2)})/\lambda}
\nonumber
\end{eqnarray}
\noindent
and the function $X(\lambda,\theta)$ is defined as
\begin{equation}
X(\lambda,\theta)=\frac{1}{L_xL_y}
\int_{-L_x/2}^{L_x/2}\!\!dx
\int_{-L_y/2}^{L_y/2}\!\!dy\,
e^{-[A_1\cos(2\pi x/\Lambda)-
A_2\cos(2\pi x^{\prime}/\Lambda)]/\lambda}.
\label{eq15}
\end{equation}

\begin{figure}[b]
\vspace*{-15cm}
\centerline{\hspace*{3cm}
\includegraphics{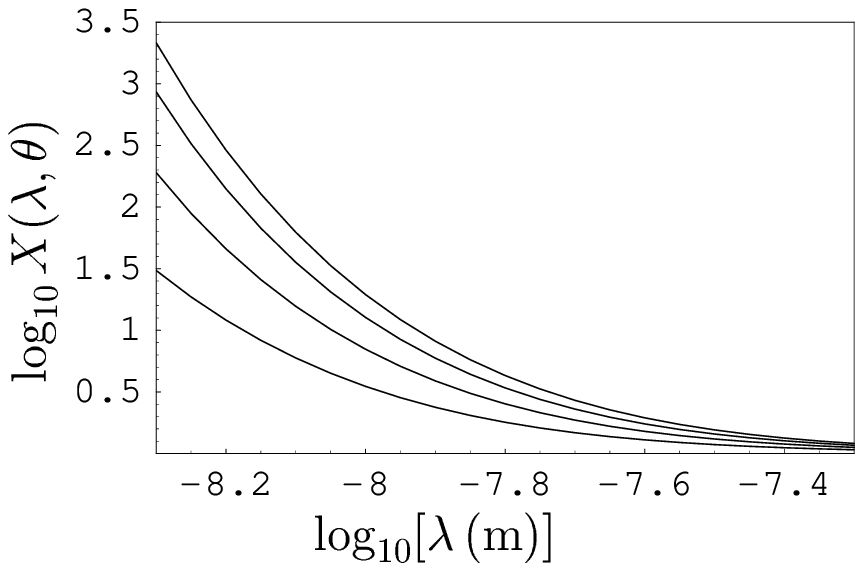}
}
\vspace*{-7cm}
\caption{\label{fg3}
The quantity $X(\lambda,\theta)$ defined in Eq.~(\ref{eq15})
is plotted by the solid lines as a function of $\lambda$ at
$\theta=0^{\circ}$, $1.2^{\circ}$, $1.8^{\circ}$, and
$2.4^{\circ}$ from bottom to top, respectively.
}
\end{figure}

For zero angle between corrugations at both surfaces ($\theta=0$)
one arrives to a more simple representation
\begin{equation}
X(\lambda,0)=\frac{1}{L_x}
\int_{-L_x/2}^{L_x/2}\!\!dx
e^{-[(A_1-A_2)\cos(2\pi x/\Lambda)]/\lambda}.
\label{eq16}
\end{equation}
\noindent
The integral in Eq.~(\ref{eq16}) can be evaluated analytically
using the formula 2.5.10(3) in Ref.~\cite{63} if there is
an integer $n$ such that $n\Lambda=L_x$. In this case
\begin{equation}
X(\lambda,0)=I_0\left(\frac{A_1-A_2}{\lambda}\right),
\label{eq17}
\end{equation}
\noindent
where $I_0(z)$ is the Bessel function of imaginary argument.
If $n\Lambda+\eta=L_x$ where $0<\eta<\Lambda$,
Eq.~(\ref{eq17}) is satisfied only approximately.
If in the interaction region of our interest (see
Fig.~\ref{fg4} below) it occurs
$(A_1-A_2)/\lambda\gg 1$,
the maximum error arising from the use of Eq.~(\ref{eq17})
achieves 5\%. In the case $(A_1-A_2)/\lambda\sim 1$
this error is equal to $\approx 2$\%.
In the general case of an arbitrary $\theta$ the quantity
$X(\lambda,\theta)$ can be computed numerically.
In Fig.~\ref{fg3} the computational results are plotted by the
solid lines as functions of $\lambda$ at
$\theta=0^{\circ}$, $1.2^{\circ}$, $1.8^{\circ}$, and
$2.4^{\circ}$ used in the experiment of Ref.~\cite{30} from
bottom to top, respectively, to a logarithmic scale.

The measurement data of Ref.~\cite{30} for the normal
Casimir force between
corrugated surfaces were compared with the results of numerical
computations based on the derivative expansion approach
\cite{31,32,33} and a good agreement was found
 within the limits
of the experimental errors $\Delta_{F_C}(a)$ determined at the
67\% confidence level (minor disagreement at the shortest
separations in Fig.~3 of Ref.~\cite{30} comes from the use
of an oscillator model in place of the optical data for the complex
index of refraction).
Then the constraints on the parameters
$\lambda$ and $\alpha$ of the corrections to Newton's law were
found from the inequality
\begin{equation}
|F_{\rm Yu}(a)|\leq \Delta_{F_C}(a),
\label{eq18}
\end{equation}
\noindent
where $F_{\rm Yu}(a)$ is given by Eq.~(\ref{eq13}) with the
notations in  Eqs.~(\ref{eq10}), (\ref{eq14}) and (\ref{eq15}).
We have numerically analyzed Eq.~(\ref{eq18}) at different
separations $a$ and with different values of the angle $\theta$
between corrugations. The strongest constraints were obtained
at the shortest separation $a=127\,$nm where
$\Delta_{F_C}=0.94\,$pN.
They are shown by the solid lines in
Figs.~\ref{fg4}(a)--\ref{fg4}(d)
at the values of
$\theta=0^{\circ}$, $1.2^{\circ}$, $1.8^{\circ}$, and
$2.4^{\circ}$, respectively. For comparison purposes, the dashed
lines 1 and 2 in Figs.~\ref{fg4}(a)--\ref{fg4}(d)
show the strongest constraints
obtained earlier \cite{26} within this interaction region from
measurements of the lateral Casimir force between sinusoidally
corrugated surfaces \cite{61,62} and from
effective measurements of the Casimir pressure between metallic
plates by means of a micromachined oscillator \cite{36,37}.
As can be seen in Fig.~\ref{fg4}, at any $\theta$ measurements
of the normal Casimir force between sinusoidally
corrugated surfaces result in stronger constraints within
some interaction region than were known so far.
Thus at $\theta=0^{\circ}$
the strengthening of previously available constraints up to
a factor 1.8 holds within the interaction region
$14.3\,\mbox{nm}\leq\lambda\leq 19.5\,$nm with the largest
strengthening achieved at $\lambda= 17.2\,$nm
[see Fig.~\ref{fg4}(a)].
At $\theta=1.2^{\circ}$ and $1.8^{\circ}$
the strengthening up to factors 2.8 and 3.5 occurs for
$13.8\,\mbox{nm}\leq\lambda\leq 25.1\,$nm  and
$12.9\,\mbox{nm}\leq\lambda\leq 27.5\,$nm , respectively.
The maximum strengthening up to a factor 4 (achieved
 at $\lambda= 17.2\,$nm) within the interaction region
$11.6\,\mbox{nm}\leq\lambda\leq 29.2\,$nm
takes place at the angle between corrugations
$\theta=2.4^{\circ}$.

The obtained stronger constraints following from measurements
of the normal Casimir force between sinusoidally
corrugated surfaces can be further strengthened at the
expense of some modification of the experimental setup.
Thus, it would be useful to switch from a static AFM mode
used in this experiment to the dynamic mode used in
Refs.~\cite{28,35,42}. This results in a higher experimental
precision though makes it necessary to perform measurements at
larger separation distances. As an example, we calculate
the prospective constraints on $\lambda,\alpha$ which can
be obtained from dynamic  measurements
of the gradient of the Casimir force between
corrugated surfaces at $a=170\,$nm. In so doing we assume
that the total experimental error  obtainable at this
experiment is $\Delta_{F_C^{\prime}}=0.62\,\mu$N/m.
For the sake of simplicity we consider the case
$\theta=0^{\circ}$ which does not lead to the maximum
strengthening of the respective constraints. Then from
Eq.~(\ref{eq13}) one obtains
\begin{equation}
\frac{\partial F_{\rm Yu}^{\rm corr}(a)}{\partial a}=
4\pi^2G\alpha\lambda^2e^{-a/\lambda}X^{(1)}(\lambda)
\tilde{X}^{(2)}(\lambda)X(\lambda,0).
\label{eq19}
\end{equation}
\noindent
Substituting Eq.~(\ref{eq19}) in the left-hand side of
 Eq.~(\ref{eq7}) adapted for the case of corrugated surfaces,
we arrive at the prospective constraints shown by the dotted
line in Fig.~\ref{fg5}. In the same figure the strongest
constraints obtained \cite{26} from
measurements of the lateral Casimir force between sinusoidally
corrugated surfaces \cite{61,62}, from
effective measurements of the Casimir pressure between metallic
plates by means of a micromachined oscillator \cite{36,37},
and from the Casimir-less experiment \cite{47} are indicated
by the dashed lines 1, 2, and 3, respectively.
As can be seen in Fig.~\ref{fg5}, the prospective constraints
shown by the dotted line are stronger than the strongest
current constraints over a wide interaction region from 12 to
160\,nm. At the moment three different experiments are used
to constrain the Yukawa-type
corrections to Newton's law
within this interaction region.
The maximum strengthening up to a factor of 12.6 occurs
at $\lambda= 17.2\,$nm.

\section{Conclusions and discussion}

In the foregoing we have obtained constraints on the parameters
of Yukawa-type corrections to Newton's law of gravitation
following from two recent experiments on measuring the Casimir
interaction. Each of these experiments is of particular
interest, as compared with all previous work in the field.
The experiment of Ref.~\cite{28} pioneered measuring the gradient
of the Casimir force between two magnetic surfaces and
confirmed the predictions of the Lifshitz theory combined with
the plasma model approach. In this way it was demonstrated that
magnetic properties of the material boundaries influence the
Casimir force. The outstanding property of magnetic materials
is that the force gradients predicted by the Drude model approach
are larger than those predicted by the plasma model approach
(just opposite to the case of nonmagnetic metals).
Thus, it was confirmed that such a widely discussed systematic
effect as the patch potentials cannot be used for the
reconciliation of the measurement data with the Drude model
approach leading to further support of constraints on
non-Newtonian gravity  obtained from the measure of agreement
between experiment and theory.
Although constraints following from the experiment with
magnetic surfaces are slightly weaker than the previously
known ones (this is due to smaller density of Ni as compared
to Au), the increased reliability can be considered as an
advantage.

The experiment of Ref.~\cite{30} pioneered measurements of
the normal Casimir force between metallized sinusoidally
corrugated surfaces at various angles between corrugations.
It was demonstrated that the Casimir force depends on these
angles in accordance with theory using
the derivative expansion. We have calculated the Yukawa-type
force in the experimental configuration with corrugated
surfaces and obtained the respective constraints on its
parameters. It was shown that the strength of constraints
increases with increasing angle between corrugations.
The maximum strengthening up to a factor of 4,
as compared to the strongest previously known constraints,
was shown to occur within the interaction range from
11.6 to 29.2\,nm.
We have also proposed some modification in the measurement
scheme allowing strengthening of the previously known
constraints up to a factor of 12.6 within a wide
interaction region presently covered  using the results
of three different experiments.
This means that measurements of the Casimir interaction
retain considerable potential for further strengthening
of constraints on the Yukawa-type corrections to Newton's
gravitational law in submicrometer interaction region.

\section*{Acknowledgments}
This work was supported by the NSF Grant
No.~PHY0970161 and DOE grant DEF010204ER46131 (U.M.).

\begin{figure}[b]
\vspace*{-1cm}
\centerline{\hspace*{1cm}
\includegraphics{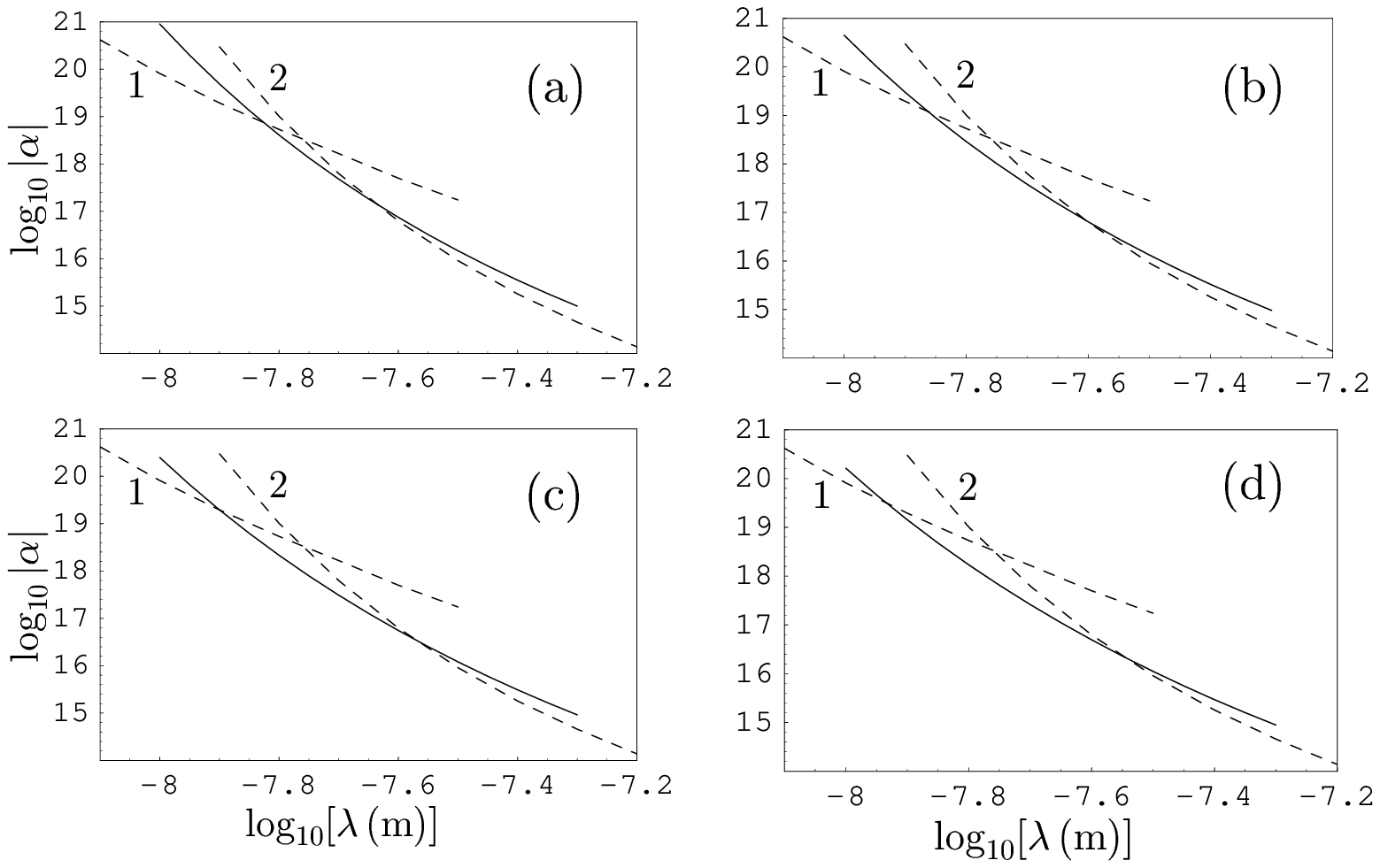}
}
\vspace*{-16cm}
\caption{\label{fg4}
Constraints on the parameters of Yukawa-type corrections to
Newton's gravitational law obtained in this work
(solid line), in Ref.~\cite{26} from
measurements of the lateral Casimir force between sinusoidally
corrugated surfaces \cite{61,62} (dashed line 1), and from
effective measurements of the Casimir pressure between metallic
plates by means of a micromachined oscillator \cite{36,37}
(dashed line 2). The angle between the axes of corrugations is
equal to (a) $\theta=0^{\circ}$, (b) $\theta=1.2^{\circ}$,
(c) $\theta=1.8^{\circ}$, and (d) $\theta=2.4^{\circ}$.
}
\end{figure}
\begin{figure}[b]
\vspace*{-13cm}
\centerline{\hspace*{3cm}
\includegraphics{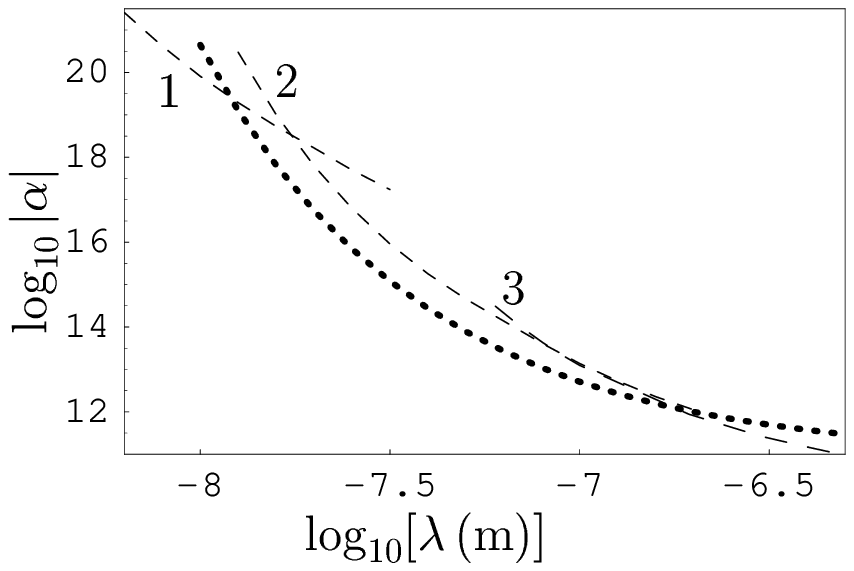}
}
\vspace*{-7cm}
\caption{\label{fg5}
Prospective
constraints on the parameters of Yukawa-type corrections to
Newton's gravitational law which can be obtained from
dynamic measurement of the gradient of the Casimir force
between sinusoidally corrugated surfaces are shown by the
dotted line. For comparison purposes the dashed lines 1, 2,
and 3  indicate the strongest current constraints obtained
 in Ref.~\cite{26} from
measurements of the lateral Casimir force between sinusoidally
corrugated surfaces \cite{61,62}, from
effective measurements of the Casimir pressure between metallic
plates by means of a micromachined oscillator \cite{36,37},
 and in Ref.~\cite{47} from the Casimir-less experiment,
 respectively.
}
\end{figure}
\end{document}